\documentclass[a4paper,11pt]{article}

\usepackage[T1]{fontenc}
\usepackage{avant}

\usepackage{multirow}

\usepackage{amsmath}
\usepackage{amssymb}
\usepackage{amsthm}
\usepackage{epsfig}
\usepackage{natbib}
\usepackage{lineno}
\usepackage{float}
\usepackage{amsthm}

\nocite{*}

\usepackage{amsfonts}
\newtheorem{definition}{Definition}
\newtheorem{theorem}{Theorem}
\newtheorem{lemma}{Lemma}

\usepackage{hyperref}

\usepackage{authblk}

\begin{document}


\title{Repulsive Mixtures}

\author[1]{\rm Francesca Petralia}
\author[2]{\rm Vinayak Rao}

\author[1]{\rm David B. Dunson}
\affil[1]{Department of Statistical Science, Box 90251, Duke University, Durham, North Carolina 27708, U.S.A. }
\affil[2]{Gatsby Computational Neuroscience Unit, University College London, London WC1N3AR, United Kingdom }

\date{}
\maketitle

\begin{abstract}
Discrete mixture models are routinely used for density estimation and clustering.  While conducting inferences on the cluster-specific parameters, current frequentist and Bayesian methods often encounter problems when clusters are placed too close together to be scientifically meaningful.  Current Bayesian practice generates component-specific parameters independently from a common prior, which tends to favor similar components and often leads to substantial probability assigned to redundant components that are not needed to fit the data.  As an alternative, we propose to generate components from a repulsive process, which leads to fewer, better separated and more interpretable clusters.  We characterize this repulsive prior theoretically and propose a Markov chain Monte Carlo sampling algorithm for posterior computation.  The methods are illustrated using simulated data as well as real datasets.
\end{abstract}
\section{ Introduction}
Finite mixture models characterize the density of $y \in \mathcal{Y} \subset \Re^m$ as
\begin{equation}
f(y) = \sum_{h=1}^k p_h \phi( y; \gamma_h), \label{eq:mix}
\end{equation}
where $p = (p_1,\ldots,p_k)^T$ is a vector of probabilities summing to one, and 
$\phi(\cdot; \gamma)$ is a kernel depending on parameters $\gamma \in \Gamma$, which may consist
of location and scale parameters.  There is a rich literature on inference for finite
mixture models from both a frequentist (\cite{unsuplearning}; \cite{finiteEM}) and Bayesian \citep{reversible} perspective.

\indent In analyses of finite mixture models, a common concern is over-fitting in which {\em redundant} mixture components having similar locations and scales are introduced.  Over-fitting can have an adverse impact on density estimation, since this leads to an unnecessarily complex model. 
Another common goal of finite mixture modeling is clustering \citep{raftery2}, and having components with similar locations, leads to overlapping kernels and lack of interpretability.  Introducing kernels with similar locations but different scales may be necessary to fit heavy-tailed and skewed densities, and hence low separation in clustering and over-fitting are distinct problems.  This article develops a repulsive mixture modeling approach which can be applied to both these problems.  

\indent Recently, \cite{Rousseau} studied the asymptotic behavior of the posterior distribution in 
over-fitted Bayesian mixture models having more components than needed.  They showed that a carefully 
chosen prior will lead to asymptotic emptying of the redundant components.  
However, several challenging practical issues arise.  For their prior and in standard Bayesian practice, one assumes that 
 $\gamma_h \sim P_0$ independently {\em a priori}.
For example, if we consider a finite location-scale mixture of multivariate Gaussians, one may choose $P_0$ to be
multivariate Gaussian-inverse Wishart.  
However, the behavior of the posterior can be sensitive to $P_0$ for finite samples, with higher variance $P_0$ 
favoring allocation to fewer clusters. In addition, drawing the component-specific parameters from a common prior tends to 
favor components located close together unless the variance is high.

\indent Regardless of the specific $P_0$ chosen, for small to moderate sample sizes, the weight assigned to redundant components 
is often substantial.  This can be attributed to identifiability problems that arise from a difficulty in distinguishing between models that partition each of a small number of well separated components into a number of essentially identical components.  This issue leads to substantial uncertainty in clustering and estimation of the number of components, and is not specific to over-fitted mixture models; similar behavior occurs in placing a prior on $k$ or using a nonparametric Bayes approach such as the Dirichlet process. 

\indent The problem of separating components has been studied for Gaussian mixture models (\cite{Dasgupta}; \cite{DasguptaSchulman}). Two Gaussians can be separated by placing an arbitrarily chosen lower bound on the distance between their means. Separated Gaussians have been mainly utilized to speed up convergence of the Expectation-Maximization (EM) algorithm. In choosing a minimal separation level, it is not clear how to obtain a good compromise between values that are too low to solve the problem and ones that are so large that one obtains a poor fit. As an alternative, we propose a repulsive prior that discourages closeness among component-specific parameters without a hard constraint.

\indent In contrast to the vast majority of the recent Bayesian literature on discrete mixture models, instead of drawing the component-specific parameters $\{ \gamma_h \}$ independently from a common prior $P_0$, we propose a joint prior for $\{ \gamma_1,\ldots,\gamma_k \}$ that is chosen to assign low density to $\gamma_h$'s located close together.  We consider two types of repulsive priors, (i) priors guarding against over-fitting by penalizing redundant kernels having close to identical locations and scales and case (ii) priors discouraging closeness in only the locations to favor well separated clusters.


\section{ Bayesian Repulsive Mixture Models}

\subsection{Background on Bayesian mixture modeling}

Considering the finite mixture model in expression (\ref{eq:mix}), a Bayesian specification is completed by choosing priors for the number of components $k$, the probability weights $p$, and the component-specific parameters $\gamma = (\gamma_1, \ldots, \gamma_k)^T$.  Typically, $k$ is assigned a Poisson or multinomial prior, $p$ a  $Dirichlet(\alpha)$  prior with $\alpha=(\alpha_1, \ldots, \alpha_k)^T$, and $\gamma_h \sim P_0$ independently, with $P_0$ often chosen to be conjugate to the kernel $\phi$.  Posterior computation can proceed via a reversible jump Markov chain Monte Carlo algorithm involving moves for adding or deleting mixture components.  Unfortunately, in making a $k \to k+1$ change in model dimension, efficient moves critically depend on the choice of proposal density. 

 \indent It has become popular to use over-fitted mixture models in which $k$ is chosen as a conservative upper bound on the number of components under the expectation that only relatively few of the components will be occupied by subjects in the sample.  
 As motivated in \cite{IshwaranZarepour}, simply letting $\alpha_h = c/k$ for $h=1,\ldots, k$ and a constant $c>0$ leads to an approximation to a Dirichlet process mixture model for the density of $y$, which is obtained in the limit as $k$ approaches infinity.  An alternative finite approximation to a Dirichlet process mixture is obtained by truncating the stick-breaking representation of Sethuraman (1994), leading to a similarly simple Gibbs sampling algorithm \citep{IshwaranJames}. These approaches are now used routinely in practice.

\subsection{ Repulsive densities}

We seek a prior on the component parameters in (\ref{eq:mix}) that automatically favors spread out components near the support of the data.  Instead of generating the atoms $\gamma_h$ independently from $P_0$, one could generate them from a repulsive process that automatically pushes the atoms apart.  This idea is conceptually related to the literature on repulsive point processes \citep{huber}.  In the spatial statistics literature, a variety of repulsive processes have been proposed.  One such model assumes that points are clustered spatially, with the vector of cluster centers $\gamma$ having a Strauss density \citep{StraussForClustering}, that is $p(k,\gamma) \propto \beta^k \rho^{r(\gamma)}$ where $k$ is the number of clusters, $\beta>0$, $0 < \rho \leq 1$ and $r(\gamma)$ is the number of pairwise centers that lie within a pre-specified distance $r$ of each other. A possibly unappealing feature is that repulsion is not directly dependent on the pairwise distances between the clusters.  We propose an alternative class of priors, which smoothly push apart components based on their pairwise distances.
\begin{definition}\label{def1} A density $h(\gamma)$ is repulsive if for any $\delta>0$ there is a corresponding $\epsilon>0$ such that $h(\gamma)<\delta$ for all $\gamma \in \Gamma \setminus G_{\epsilon}$, where $G_{\epsilon} = \{ \gamma : d(\gamma_s,\gamma_j)>\epsilon; s=1, \ldots, k; j<s \}$ and $d$ is a distance.
 \end{definition}
\indent We consider two special cases (i) $d(\gamma_s,\gamma_j)$ is the distance between the $s$th and $j$th kernel, (ii) $d(\gamma_s,\gamma_j)$ is the distance between sub-vectors of $\gamma_s$ and $\gamma_j$ corresponding to only locations.  Priors following definition  \ref{def1}(i) limit over-fitting in density estimation, while priors following definition \ref{def1}(ii)  favor well-separated clusters. 

\indent 
As a convenient class of repulsive priors which smoothly push components apart, we propose
 \begin{equation}\pi(\gamma) = c_1\left( \prod_{j=1}^k g_0(\gamma_j)\right) h(\gamma), \label{joint1} \end{equation}
with $c_1$ being a normalizing constant that can be intractable to calculate. The dependence of $c_1$ on $k$ leads to complications in estimating $k$ that motivate the use of an over-specified mixture that treats $k$ as an upper bound on the number of components. 
The proposed prior is closely related to a class of point processes from the statistical physics and spatial statistics literature called
Gibbs processes \citep{DalVer2008a}. We assume $g_0: \Gamma \to \Re_+$ and $h: \Gamma^k \to [0,\infty)$ are continuous with respect to Lesbesgue measure, and $h$ is bounded  above by a positive constant $c_2$ and is repulsive according to definition \ref{def1} with $d$ differing across cases.  It follows that density $\pi$ defined in (\ref{joint1}) is also repulsive.  For location-scale kernels, let $\gamma_j=(\mu_j,\Sigma_j)$ and $g_0(\mu_j,\Sigma_j)=\xi(\mu_j) \psi(\Sigma_j)$ with $\mu_j$ and $\Sigma_j$ being respectively the location and the scale parameters. A special hardcore repulsion is produced if the repulsion function is zero when at least one pairwise distance is smaller than a pre-specified threshold. Such a density implies choosing a minimal separation level between the atoms. 

\begin{figure}
\centering
\includegraphics[width=90mm]{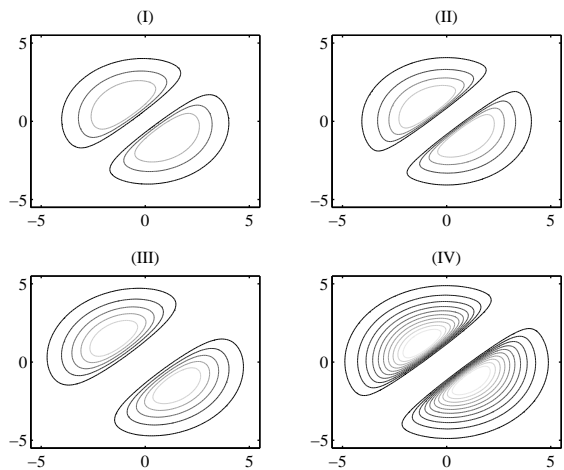}
\caption{Contour plots of the repulsive prior $\pi(\gamma_1,\gamma_2)$  satisfying definition \ref{def1}(ii) under $(\ref{joint1})$ either $(\ref{repul2}) $ or $(\ref{repul}) $ and $(\ref{gfunction})$ with hyperparameters $(\tau,\nu)$ equal to $(I) (1,2), (II) (1,4), (III) (5,2)$ and $(IV) (5,4)$} \label{prior_nutau}
\end{figure}

\indent We avoid hard separation thresholds by considering repulsive priors that smoothly push components apart. In particular, we propose two repulsion functions defined as\\
\begin{minipage}{0.5\linewidth} \begin{equation}h(\gamma)=\prod_{ \{(s, j) \in A\} } g\{d(\gamma_s, \gamma_j)\}  \label{repul2}\end{equation}
\end{minipage} 
\begin{minipage}{0.5\linewidth} 
\begin{equation}h(\gamma)= \min_{ \{(s, j) \in A\} } g\{d(\gamma_s, \gamma_j)\}  \label{repul}\end{equation}\end{minipage}
with $A = \{ (s,j): s=1,\ldots, k; j < s \}$ and $g:\Re_+ \to [0,M]$ a strictly monotone differentiable function with $g(0)=0$, $g(x)>0$ for all $x>0$ and $M<\infty$.  It is straightforward to show that $h$ in (\ref{repul2}) and (\ref{repul}) is integrable and satisfies definition \ref{def1}.  The two alternative repulsion functions differ in their dependence on the relative distances between components, with all the pairwise distances playing a role in (\ref{repul2}), while (\ref{repul}) only depends on the minimal separation. A flexible choice of $g$ corresponds to
\begin{eqnarray}
g\{ d(\gamma_s,\gamma_j) \} = \exp\big[ - \tau \{d(\gamma_s,\gamma_j)\}^{-\nu} \big], \label{gfunction}
\end{eqnarray}
where $\tau>0$ is a scale parameter and $\nu$ is a positive integer controlling the rate at which $g$ approaches zero as $d(\gamma_s,\gamma_j)$ decreases. 
Figure \ref{prior_nutau} shows contour plots of the prior $\pi(\gamma_1,\gamma_2)$ defined as (\ref{joint1}) and satisfying definition \ref{def1}(ii) with $\gamma_1, \gamma_2 \in \mathbb{R}$, $d$ the Euclidean distance, $g_0$ the standard normal density, the repulsive function defined as (\ref{repul2}) or (\ref{repul}) and $g$ defined as (\ref{gfunction})  for different values of $(\tau,\nu)$.  As $\tau$ and $\nu$ increase, the prior increasingly favors well separated components. 

 \subsection{Theoretical properties}\label{theory}

In this section, theoretical properties of the proposed prior are considered under definition \ref{def1}(ii) for simplicity, though all results can be modified to accommodate definition \ref{def1}(i).  For some results, the kernel will be assumed to depend only on location parameters, while for others on both location and scale parameters.  Let $\Pi$ be the prior induced on $\cup_{j=1}^{\infty} \mathcal{F}_k$, where $\mathcal{F}_k$ is the space of all distributions defined as (\ref{eq:mix}). Let $\|\cdot\|_1$ denote the $L_1$ norm and $KL(f_0,f) = \int f_0 \log(f_0/f)$ refer to the Kullback-Leibler (K-L) divergence between $f_0$ and $f$. Density $f_0$ belongs to the K-L support of the prior $\Pi$ if $\Pi\{ f: KL(f_0,f)< \epsilon \}>0$ for all $\epsilon>0$. Let the true density $f_0:\Re^m \to \Re_+$ be defined as $f_0=\sum_{h=1}^{k_0} p_{0h} \phi(\gamma_{0h})$ with $\gamma_{0h}\in \Gamma$ and $\gamma_{0j}$s such that there exists an $\epsilon_1>0$ such that $\min_{\{(s,j): s<j\}} d(\gamma_{0s},\gamma_{0j})\geq \epsilon_1$, $d$ being the Euclidean distance of sub-vectors of $\gamma_{0j}$ and $\gamma_{0s}$ corresponding to only locations. Let $f=\sum_{h=1}^{k}p_h \phi(\gamma_h)$ with $\gamma_h \in \Gamma$. Let $\gamma \sim \pi$ and $\pi$ satisfy definition \ref{def1}(ii). Let $p \sim \lambda$ with $\lambda=Dirichlet(\alpha)$ and $k \sim \vartheta$ with $\vartheta(k=k_0)>0$. Let $\theta=(p,\gamma)$. These assumptions on $f_0$ and $f$ will be referred to as condition B0.  The next lemma provides sufficient conditions under which the true density is in the K-L support of the prior for location kernels.

\begin{lemma}\label{kl} Assume condition B0 is satisfied with $m=1$. Let $D_0$ be a compact set containing location parameters $(\gamma_{01}, \ldots, \gamma_{0k_0})$.  Let $\phi$ and $\pi$ satisfy the following conditions:

 \indent A1. for any $y \in \mathcal{Y}$, the map $\gamma \to \phi(y; \gamma)$ is uniformly continuous
 
\indent A2. for any $y \in \mathcal{Y}$, $\phi(y; \gamma)$ is bounded above by a constant

\indent A3. $\int f_0 \left|\log \left\lbrace \sup_{\gamma \in D_{0}} \phi(\gamma)\right\rbrace -\log\left\lbrace\inf_{\gamma \in D_{0}} \phi(\gamma)\right\rbrace\right| < \infty$

\indent A4. $\pi$ is continuous with respect to Lebesgue measure and for any vector $x \in \Gamma^k$  \indent \indent with $\min_{\{(s,j): s<j\}} d(x_{s},x_{j})\geq \upsilon$ for $\upsilon>0$ there is a $\delta>0$ such that $\pi(\gamma)>0$ for all $\gamma$ \indent \indent satisfying $|| \gamma-x ||_1 < \delta$
 
Then $f_0$ is in the K-L support of the prior $\Pi$.
\end{lemma} 

\begin{lemma}\label{A4}
The repulsive density in (\ref{joint1}) with $h$ defined as either (\ref{repul2}) or (\ref{repul}) satisfies condition A4 in lemma \ref{kl}.
\end{lemma}

The next lemma formalizes the posterior rate of concentration for univariate location mixtures of Gaussians.

\begin{lemma}\label{contraction:rate}
Let condition B0 be satisfied, let $m=1$ and $\phi$ be the normal kernel depending on a location parameter $\mu$ and a scale parameter $\sigma$. Assume that condition $(i), (ii)$ and $(iii)$ of theorem 3.1 in \cite{Scricciolo2011} and assumption A4 in lemma \ref{kl} are satisfied. Furthermore, assume that

\indent C1) the joint density $\pi$ leads to exchangeable random variables and for all $k$ the marginal \indent density of $\mu_1$ satisfies $\pi_m(|\mu_1|\geq t) \lesssim \exp\left(-q_1 t^2\right)$ for a given $q_1>0$

\indent C2) there are constants $u_1, u_2, u_3 >0$, possibly depending on $f_0$, such that for any $\epsilon \leq u_3$ \[\pi(||\mu-\mu_0 ||_1\leq \epsilon) \geq u_1 \exp(-u_2 k_0 \log(1/\epsilon))\] 

Then the posterior rate of convergence relative to the $L_1$ metric is $\epsilon_n= n^{-1/2} \log n$.\end{lemma}
Lemma \ref{contraction:rate} is basically a modification of theorem 3.1 in \cite{Scricciolo2011} to our proposed repulsive mixture model. Lemma \ref{lemma_condition} gives sufficient conditions for $\pi$ to satisfy condition C1 and C2 in lemma \ref{contraction:rate}.

\begin{lemma} \label{lemma_condition}
Let $\pi$ be defined as (\ref{joint1}) and $h$ be defined as either (\ref{repul2}) or (\ref{repul}), then $\pi$ satisfies condition C2 in lemma \ref{contraction:rate}. Furthermore, if for a positive constant $n_1$ the function $\xi$ satisfies $\xi(|x|\geq t)\lesssim \exp(-n_1 t^2)$, $\pi$ satisfies condition C1 in lemma \ref{contraction:rate}.
\end{lemma}

 As motivated above, when the number of mixture components is chosen to be conservatively large, it is appealing for the posterior distribution of the weights of the extra components to be concentrated near zero.  Theorem \ref{weights:theorem} formalizes the rate of concentration with increasing sample size $n$.  One of the main assumptions required in theorem  \ref{weights:theorem} is that the posterior rate of convergence relative to the $L_1$ metric is $\delta_n=n^{-1/2}(\log n)^q$ with $q\geq 0$. We provided the contraction rate, under the proposed prior specification and  univariate Gaussian kernel, in lemma \ref{contraction:rate}. However, theorem \ref{weights:theorem} is a more general statement and it applies to  multivariate mixture density of any kernel. 

\begin{theorem}\label{weights:theorem}
Let assumptions $B0-B5$ be satisfied. Let $\pi$ be defined as (\ref{joint1}) and $h$ be defined as either (\ref{repul2}) or (\ref{repul}). If $\bar{\alpha}=\max(\alpha_1, \ldots, \alpha_k)<m/2$ and for positive constants $r_1, r_2, r_3$ the function $g$ satisfies $g(x) \leq r_1 x^{r_2}$ for $0\leq x<r_3$  then
\[\lim_{M\to\infty}\limsup_{n \to \infty} E^0_n \left[ P \left\lbrace \min_{\{\iota \in S_k\}} \left( \sum_{i=k_0+1}^{k} p_{\iota(i)}\right)> M n^{-1/2}  (\log n)^{q(1+s(k_0,\alpha)/s_{r_2})} \right\rbrace\right] = 0\]
with $s(k_0,\alpha)=k_0-1+m k_0+\bar{\alpha}(k-k_0)$, $s_{r_2}=r_2+m/2-\bar{\alpha}$ and $S_k$ the set of all possible permutations of $\{1, \ldots, k\}$.
\end{theorem}

Theorem \ref{weights:theorem} is a modification of theorem 1 in Rousseau and Mengersen (2011) to our proposed repulsive mixture model. 
Theorem \ref{weights:theorem} implies that the posterior expectation of weights of the extra components is of order $O(n^{-1/2} (\log n)^{q(1+s(k_0,\alpha)/s_{r_2})})$.  When $g$ is defined as (\ref{gfunction}), parameters $r_1$ and $r_2$ can be chosen such that $r_1=\tau$ and $r_2=\nu$. 

\indent When the number of components is unknown, with only an upper bound known, the posterior rate of convergence is equivalent to the parametric rate $n^{-1/2}$ \citep{ratepost}. In this case, the rate in theorem \ref{weights:theorem} is $n^{-1/2}$ under usual priors or our repulsive prior.  However, in our experience using usual priors, the sum of the extra components can be substantial in small to moderate sample sizes, and often has high variability.  As we show in Section \ref{simsection}, for repulsive priors the sum of the extra component weights is close to zero and has small variance for small as well as large sample sizes.  When an upper bound on the number of components is unknown, the posterior rate of concentration is $n^{-1/2} (\log n)^q$ with $q>0$. In this case, according to theorem \ref{weights:theorem}, using our prior specification the logarithmic factor in theorem 1 of \cite{Rousseau} can be improved.

\section{Parameter Calibration and Posterior Computation}

An important issue in implementing repulsive mixture models is elicitation of the repulsion hyper-parameters $(\tau,\nu)$.  Although a variety of strategies can be considered, we propose a simple approach that can be used to obtain a default hyper-parameter choice in general applications.  In case (i) we choose $d(\cdot,\cdot)$ as the symmetric Kullback-Leibler divergence defined for Gaussian kernels as
 \[ s_{12} = d(\gamma_1,\gamma_2)=tr(\Sigma_1\Sigma_2^{-1})+tr(\Sigma_1^{-1} \Sigma_2)-2m+(\mu_1-\mu_2)^T(\Sigma_1^{-1}+\Sigma_2^{-1})(\mu_1-\mu_2), \]
while in case (ii) we use the Euclidean distance between the location parameters. For both case (i) and case (ii), define $\bar{d}$ as the mean of pairwise distances between atoms, $\bar{d}=\frac{1}{n(A)}\sum_{(s,j)\in A} d(\gamma_s,\gamma_j)$ with $A=\{ (s,j): s=1, \ldots, k; j<s \}$ and $n(A)$ the cardinality of set $A$.  Let $f_1$ and $f_2$ denote the densities of $\bar{d}$ under repulsive and non-repulsive priors respectively, with $(\varrho_j,\	\varsigma_j)$ the mean and standard deviation of $f_j$ for $j=1,2$.  We choose $(\tau,\nu)$ so that $f_1$ and $f_2$ are well-separated using the following definition of separation \citep{Dasgupta}.
\begin{definition}\label{def2} Given a positive constant $c$, $f_1$ and $f_2$ are $c$-separated if $\varrho_1-\varrho_2 \geq c \max(\varsigma_1,\varsigma_2)$.
\end{definition}
We have found that $\nu=2$ and $\nu=1$ provide good default values in case (i) and (ii) respectively and we fix $\nu$ at these values in all our applications below. For a given value of $\nu$, $\tau$ is found by starting with small values, estimating the mean and variance of $\bar{d}$ through Monte Carlo draws, and incrementing $\tau$ until definition \ref{def2} is satisfied for a pre-specified $c$.  We use $c=4$ in our implementations.

\indent For posterior computation, we use a slice sampling algorithm \citep{neal}, a class of Markov chain Monte Carlo algorithms 
widely used for posterior inference in infinite mixture models \citep{slice}. Letting $g_0$ be a conjugate prior, introduce a latent variable 
$u$ which is jointly modeled with $\gamma$ through
\begin{equation*}\pi(\gamma_1, \ldots, \gamma_k, u) \propto \left(\prod_{h=1}^k g_0(\gamma_h ) \right) 1\left\{ h(\gamma_1, \ldots, \gamma_k)>u\right\}. \end{equation*}
Here $1(B)$ is the indicator function, equalling $1$ if the event $B$ occurs and $0$ otherwise. Marginalizing out $u$, we recover the original density $\pi(\gamma_1, \ldots, \gamma_k)$. For a repulsion function defined as (\ref{repul}), let $B_j \equiv \bigcap_{\{s: s\not= j\}}\left[ \gamma_j: g\{d(\gamma_s,\gamma_j)\}>u\right]$. As long as $g$ is invertible in its 
argument, the set $B_j$ can be calculated, making sampling straightforward. When the repulsion function is defined as (\ref{repul2}), one can introduce a latent variable for each
product term. Under repulsive priors satisfying definition \ref{def1}(i), the set $B_j$ might not be easy to compute. However, when covariance matrices are constrained to be diagonal, vectors $\gamma_j$s can be easily sampled element-wise. For multivariate observations, the location parameter vector can be sampled element-wise from truncated distributions. Details can be found in the supplementary materials. 

\section{Synthetic Examples} \label{simsection}

\begin{figure}
\centering
\includegraphics[width=90mm]{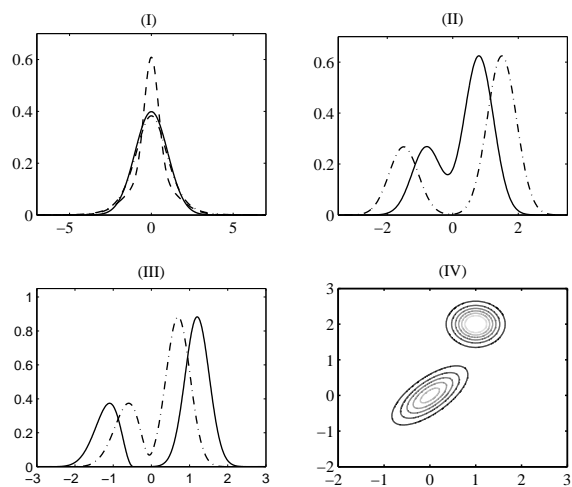}
\caption{$(I)$ Standard normal density (solid), two-component mixture of normals sharing the same location parameter (dash) and Student's t density (dash-dot), referred as $(I a, I b, I c)$, $(II)$ two-components mixture of poorly (solid) and well separated (dot-dash) Gaussian densities, referred as $(II a, II b)$, $(III)$ mixture of poorly (dot-dash) and well separated (solid) Gaussian and Pearson densities, referred as $(III a, III b)$, $(IV)$ two-components mixture of two-dimensional non-spherical Gaussians} \label{figure:datasets}
\end{figure}

Simulation examples were considered to assess the performance of the repulsive prior in density estimation, clustering and emptying of extra components.  Figure \ref{figure:datasets} plots the true densities in the various cases that we considered.  For each synthetic dataset, repulsive and non-repulsive mixture models were compared considering a fixed upper bound on the number of components; extra components should be assigned small probabilities and hence effectively excluded. The slice sampler was run for $10,000$ iterations with a burn-in of $5,000$. The chain was thinned by keeping every 10th draw. To overcome the label switching problem, the samples were post-processed following the algorithm of \cite{Stephens}. Details on parameters involved in the true densities, choice of prior distributions and methods used to compute quantities presented in this section can be found in the supplement. 

Repulsive mixtures satisfying definition \ref{def1}(i) and non-repulsive mixtures were compared. For this experiment $1,000$ draws from a standard normal density and a two component mixture of overlapping normals was considered. Both repulsive and non-repulsive mixtures were run considering six as the upper bound of the number of components. Table \ref{table:par} shows posterior summaries of parameters involved in the components with highest weights. Clearly, repulsive mixtures lead to a more parsimonious representation of the true densities and more accurate parameter estimates. The mean and standard deviation of the K-L divergence under the first data example were $(0$$\cdot$$003,0$$\cdot$$002)$ and  $(0$$\cdot$$004,0$$\cdot$$002)$ for non-repulsive and repulsive mixtures respectively; while under the second data example were $(0$$\cdot$$006,0$$\cdot$$003)$ and $(0$$\cdot$$009,0$$\cdot$$003)$ for non-repulsive and repulsive mixtures respectively. Therefore, repulsive mixtures were able to concentrate more on the reduced model while performing similarly to non-repulsive mixtures in estimating the true density.
  
Repulsive mixtures satisfying definition \ref{def1} (ii) and non-repulsive mixtures were compared to assess clustering performance. Table \ref{table:mis} shows summary statistics of the  K-L divergence, the misclassification error and the sum of extra weights under repulsive and non-repulsive mixtures with six mixture components as the upper bound. Table \ref{table:mis} shows also the misclassification error resulting from hierarchical clustering \citep{hierarchical}. In practice, observations drawn from the same mixture component were considered as belonging to the same category and for each dataset a similarity matrix was constructed. The misclassification error was established in terms of divergence between the true similarity matrix and the posterior similarity matrix. As shown in table \ref{table:mis}, the K-L divergences under repulsive and non-repulsive mixtures become more similar as the sample size increases. For smaller sample sizes, the results are more similar when components are very well separated.  Since a repulsive prior tends to discourage overlapping mixture components, a repulsive model might not estimate the density quite as accurately when a mixture of closely overlapping components is needed.  However, as the sample size increases, the fitted density approaches the true density regardless of the degree of closeness among clusters. Again, though repulsive and non-repulsive mixtures perform similarly in estimating the true density, repulsive mixtures place considerably less probability on extra components leading to more interpretable clusters. In terms of misclassification error, the repulsive model outperforms the other two approaches while, in most cases, the worst performance was obtained by the non-repulsive model. 

Potentially, one may favor fewer clusters, and hence possibly better separated clusters, by penalizing the introduction of new clusters more through modifying the precision in the Dirichlet prior for the weights; in the supplemental materials, we demonstrate that this cannot solve the problem.  
 
\begin{table}[h!]
\centering
\caption{Posterior mean and standard deviation of weights, location and scale parameters under dataset drawn from densities $(Ia, Ib)$}{
\scalebox{0.9}{
\begin{tabular}{lccccccccccccccccccc}
&\multicolumn{3}{c}{Density Ia}&\multicolumn{6}{c}{Density Ib}  \\
&\multicolumn{3}{c}{Comp 1}&\multicolumn{3}{c}{Comp 1}&\multicolumn{3}{c}{Comp 2}  \\
&$\hat{p}_1$&$\hat{\mu}_1$&$\hat{\sigma}_1$ &$\hat{p}_1$&$\hat{\mu}_1$&$\hat{\sigma}_1$&$\hat{p}_2$&$\hat{\mu}_2$&$\hat{\sigma}_2$\\
\begin{footnotesize} True \end{footnotesize}  &$1$&$0$&$1$&$0$$\cdot$$7$&$0$&$0$$\cdot$$2$&$0$$\cdot$$3$&$0$&$2$\\
\begin{footnotesize} N-R \end{footnotesize}  & $0$$\cdot$$53$&$-0$$\cdot$$01$ &$0$$\cdot$$85$&$0$$\cdot$$44$ &$0$$\cdot$$08$&$1$$\cdot$$21$&$0$$\cdot$$34$&$0$$\cdot$$12$&$1$$\cdot$$33$\\
& $$\mbox{\footnotesize$(0$$\cdot$$16)$}$$ &$$\mbox{\footnotesize$(0$$\cdot$$04)$}$$&$$\mbox{\footnotesize$(0$$\cdot$$25)$}$$&$$\mbox{\footnotesize$(0$$\cdot$$06)$}$$&$$\mbox{\footnotesize$(0$$\cdot$$10)$}$$&$$\mbox{\footnotesize$(1$$\cdot$$05)$}$$&$$\mbox{\footnotesize$(0$$\cdot$$06)$}$$ &$$\mbox{\footnotesize$(0$$\cdot$$16)$}$$&$$\mbox{\footnotesize$(1$$\cdot$$11)$}$$\\
\begin{footnotesize} R \end{footnotesize}  &$0$$\cdot$$87$&$-0$$\cdot$$00$&$0$$\cdot$$84$&$0$$\cdot$$67$&$-0$$\cdot$$02$&$0$$\cdot$$28$&$0$$\cdot$$27$&$0$$\cdot$$09$&$2$$\cdot$$36$\\
&$$\mbox{\footnotesize$(0$$\cdot$$07)$}$$ &$$\mbox{\footnotesize$(0$$\cdot$$01)$}$$&$$\mbox{\footnotesize$(0$$\cdot$$04)$}$$&$$\mbox{\footnotesize$(0$$\cdot$$05)$}$$&$$\mbox{\footnotesize$(0$$\cdot$$03)$}$$&$$\mbox{\footnotesize$(0$$\cdot$$02)$}$$&$$\mbox{\footnotesize$(0$$\cdot$$09)$}$$&$$\mbox{\footnotesize$(0$$\cdot$$23)$}$$&$$\mbox{\footnotesize$(0$$\cdot$$75)$}$$\\
\end{tabular}}}\label{table:par}
\end{table}   

\begin{small}
\begin{table}[h!]
\centering
\caption{Mean and standard deviation of K-L divergence, misclassification error and sum of extra weights resulting from non-repulsive mixture and repulsive mixture with a maximum number of clusters equal to six under different synthetic data scenarios.}{ \scalebox{0.9}{
\begin{tabular}{lccccccccccccccccccccc}
&\multicolumn{6}{c}{n=100}&\multicolumn{6}{c}{n=1000}  \\
& $$\mbox{\footnotesize$I c$}$$ & $$\mbox{\footnotesize$IIa$}$$ & $$\mbox{\footnotesize$IIb$}$$ & $$\mbox{\footnotesize$IIIa$}$$& $$\mbox{\footnotesize$IIIb$}$$&$$\mbox{\footnotesize$IV$}$$& $$\mbox{\footnotesize$I c$}$$ & $$\mbox{\footnotesize$IIa$}$$ & $$\mbox{\footnotesize$IIb$}$$ & $$\mbox{\footnotesize$IIIa$}$$& $$\mbox{\footnotesize$IIIb$}$$&$$\mbox{\footnotesize$IV$}$$\\
  \multicolumn{10}{l}{\begin{footnotesize}K-L divergence \end{footnotesize}}& \\
\begin{footnotesize} N-R \end{footnotesize} &$0$$\cdot$$05$&$0$$\cdot$$03$&$0$$\cdot$$07$&$0$$\cdot$$05$&$0$$\cdot$$08$&$0$$\cdot$$22$&$0$$\cdot$$01$&$0$$\cdot$$01$&$0$$\cdot$$01$&$0$$\cdot$$01 $&$0$$\cdot$$01 $&$0$$\cdot$$02$ \\
&$$\mbox{\footnotesize$(0$$\cdot$$03)$}$$&$$\mbox{\footnotesize$(0$$\cdot$$01)$}$$&$$\mbox{\footnotesize$(0$$\cdot$$02)$}$$&$$\mbox{\footnotesize$(0$$\cdot$$02)$}$$&$$\mbox{\footnotesize$(0$$\cdot$$03)$}$$&$$\mbox{\footnotesize$(0$$\cdot$$05)$}$$&$$\mbox{\footnotesize$(0$$\cdot$$00)$}$$ &$$\mbox{\footnotesize$(0$$\cdot$$00)$}$$ &$$\mbox{\footnotesize$(0$$\cdot$$00)$}$$&$$\mbox{\footnotesize$(0$$\cdot$$00)$}$$&$$\mbox{\footnotesize$(0$$\cdot$$00)$}$$&$$\mbox{\footnotesize$(0$$\cdot$$01)$}$$\\
\begin{footnotesize} R \end{footnotesize}&$0$$\cdot$$06$&$0$$\cdot$$05$&$0$$\cdot$$08$&$0$$\cdot$$07$&$0$$\cdot$$09$&$0$$\cdot$$28$&$ 0$$\cdot$$01$&$0$$\cdot$$01$&$0$$\cdot$$01$&$0$$\cdot$$01$&$0$$\cdot$$01$&$0$$\cdot$$03$ \\
&$$\mbox{\footnotesize$(0$$\cdot$$03)$}$$&$$\mbox{\footnotesize$(0$$\cdot$$02)$}$$&$$\mbox{\footnotesize$(0$$\cdot$$03)$}$$&$$\mbox{\footnotesize$(0$$\cdot$$03)$}$$&$$\mbox{\footnotesize$(0$$\cdot$$03)$}$$&$$\mbox{\footnotesize$(0$$\cdot$$04)$}$$&$$\mbox{\footnotesize$(0$$\cdot$$00)$}$$&$$\mbox{\footnotesize$(0$$\cdot$$00)$}$$&$$\mbox{\footnotesize$(0$$\cdot$$00)$}$$&$$\mbox{\footnotesize$(0$$\cdot$$00)$}$$&$$\mbox{\footnotesize$(0$$\cdot$$00)$}$$ &$$\mbox{\footnotesize$(0$$\cdot$$01)$}$$\\
\multicolumn{10}{l}{\begin{footnotesize}Misclassification \end{footnotesize}}& \\
\multirow{1}{*}{\begin{footnotesize} HCT \end{footnotesize}}& $0$$\cdot$$12$ &$0$$\cdot$$11$ &$0$$\cdot$$04$&$0$$\cdot$$12$&$0$$\cdot$$08$&$0$$\cdot$$21$& $0$$\cdot$$05$&$0$$\cdot$$42$&$0$$\cdot$$01$&$0$$\cdot$$42$&$0$$\cdot$$01$&$0$$\cdot$$20$\\
\begin{footnotesize} N-R \end{footnotesize} & $0$$\cdot$$69$&$ 0$$\cdot$$26$&$0$$\cdot$$06$&$0$$\cdot$$17$&$0$$\cdot$$05$&$0$$\cdot$$13 $&$0$$\cdot$$65$&$0$$\cdot$$24$&$0$$\cdot$$03$&$0$$\cdot$$14$&$0$$\cdot$$03$&$0$$\cdot$$19$\\
& $$\mbox{\footnotesize$(0$$\cdot$$10)$}$$&$$\mbox{\footnotesize$(0$$\cdot$$10)$}$$&$$\mbox{\footnotesize$(0$$\cdot$$04)$}$$&$$\mbox{\footnotesize$(0$$\cdot$$09)$}$$&$$\mbox{\footnotesize$(0$$\cdot$$06)$}$$&$$\mbox{\footnotesize$(0$$\cdot$$05)$}$$&$$\mbox{\footnotesize$(0$$\cdot$$11)$}$$&$$\mbox{\footnotesize$(0$$\cdot$$08)$}$$&$$\mbox{\footnotesize$(0$$\cdot$$04)$}$$&$$\mbox{\footnotesize$(0$$\cdot$$09)$}$$&$$\mbox{\footnotesize$(0$$\cdot$$03)$}$$&$$\mbox{\footnotesize$(0$$\cdot$$02)$}$$\\
 \begin{footnotesize} R \end{footnotesize} & $0$$\cdot$$53$&$0$$\cdot$$18$&$0$$\cdot$$01$&$0$$\cdot$$10$&$0$$\cdot$$01$&$0$$\cdot$$05$&$0$$\cdot$$46$&$0$$\cdot$$13$&$0$$\cdot$$00$&$0$$\cdot$$03$&$0$$\cdot$$00$&$0$$\cdot$$17$\\
& $$\mbox{\footnotesize$(0$$\cdot$$10)$}$$& $$\mbox{\footnotesize$(0$$\cdot$$09)$}$$&$$\mbox{\footnotesize$(0$$\cdot$$02)$}$$&$$\mbox{\footnotesize$(0$$\cdot$$05)$}$$&$$\mbox{\footnotesize$(0$$\cdot$$01)$}$$&$$\mbox{\footnotesize$(0$$\cdot$$02)$}$$&$$\mbox{\footnotesize$(0$$\cdot$$16)$}$$&$$\mbox{\footnotesize$(0$$\cdot$$04)$}$$&$$\mbox{\footnotesize$(0$$\cdot$$01)$}$$&$$\mbox{\footnotesize$(0$$\cdot$$02)$}$$&$$\mbox{\footnotesize$(0$$\cdot$$01)$}$$&$$\mbox{\footnotesize$(0$$\cdot$$01)$}$$\\
  \multicolumn{10}{l}{\begin{footnotesize}Sum of extra weights \end{footnotesize}}& \\
\begin{footnotesize} N-R \end{footnotesize} & $0$$\cdot$$30$&$0$$\cdot$$21$& $0$$\cdot$$09$&$0$$\cdot$$16$&$0$$\cdot$$07$&$0$$\cdot$$13$&$0$$\cdot$$30$&$0$$\cdot$$21$&$0$$\cdot$$03$ &$0$$\cdot$$16$&$0$$\cdot$$03$&$0$$\cdot$$29$\\
&$$\mbox{\footnotesize$(0$$\cdot$$10)$}$$&$$\mbox{\footnotesize$(0$$\cdot$$11)$}&$$\mbox{\footnotesize$(0$$\cdot$$07)$}&$$\mbox{\footnotesize$(0$$\cdot$$09)$}&$$\mbox{\footnotesize$(0$$\cdot$$07)$}&$$\mbox{\footnotesize$(0$$\cdot$$08)$}&$$\mbox{\footnotesize$(0$$\cdot$$11)$}$$&$$\mbox{\footnotesize$(0$$\cdot$$11)$}$$&$$\mbox{\footnotesize$(0$$\cdot$$04)$}$$&$$\mbox{\footnotesize$(0$$\cdot$$10)$}$$&$$\mbox{\footnotesize$(0$$\cdot$$03)$}$$&$$\mbox{\footnotesize$(0$$\cdot$$03)$}$$\\
\begin{footnotesize} R \end{footnotesize} &$ 0$$\cdot$$08$&$0$$\cdot$$08$&$ 0$$\cdot$$02$   &$0$$\cdot$$04$&$0$$\cdot$$02$&$0$$\cdot$$06$&$0$$\cdot$$10$& $0$$\cdot$$09$&$0$$\cdot$$00$&$0$$\cdot$$01$&$0$$\cdot$$00$&$0$$\cdot$$25$\\
 &$$\mbox{\footnotesize$(0$$\cdot$$05)$}$$&$$\mbox{\footnotesize$(0$$\cdot$$07)$}&$$\mbox{\footnotesize$(0$$\cdot$$02)$}&$$\mbox{\footnotesize$(0$$\cdot$$05)$}&$$\mbox{\footnotesize$(0$$\cdot$$02)$}&$$\mbox{\footnotesize$(0$$\cdot$$03)$}&$$\mbox{\footnotesize$(0$$\cdot$$04)$}$$&$$\mbox{\footnotesize$(0$$\cdot$$06)$}$$&$$\mbox{\footnotesize$(0$$\cdot$$00)$}$$&$$\mbox{\footnotesize$(0$$\cdot$$01)$}$$&$$\mbox{\footnotesize$(0$$\cdot$$00)$}$$&$$\mbox{\footnotesize$(0$$\cdot$$03)$}$$\\
  \end{tabular}}} \label{table:mis}
\end{table}    
\end{small}
\section{Real data}
We tested the performance of our proposed prior specification on three real datasets. The first involves $82$ measurements of the velocities in km/s of galaxies 
diverging from our own (\cite{escobar}, \cite{reversible}), the second consists of the acidity index measured in a sample of $155$ lakes in north central Wisconsin 
(\cite{reversible}), and the third consists of 150 observations from three different species of iris each with four measurements \citep{clusterCV}. 

For the first two datasets, a repulsive mixture satisfying definition \ref{def1}(i) was considered and a five-component mixture model was fit while for the third dataset a repulsive mixture satisfying definition \ref{def1}(ii) was considered and both six components and ten components were considered as the upper bound. The same prior specification, Markov chain Monte Carlo sampler, and 
relabeling technique as in section \ref{simsection} were utilized. 
\begin{figure}
\centering
 \includegraphics[width=70mm]{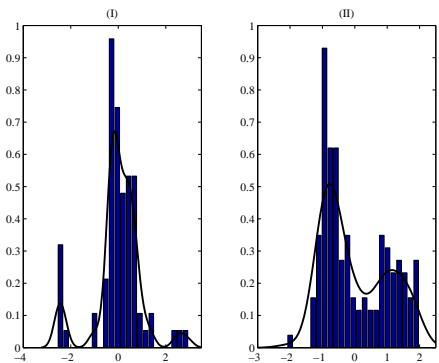}
\caption{Histogram of galaxy data (I) and acidity data (II) overlaid with a nonparametric density estimate using Gaussian kernel density estimation} \label{data:real}
\end{figure}

\begin{figure}
\centering
\includegraphics[width=100mm]{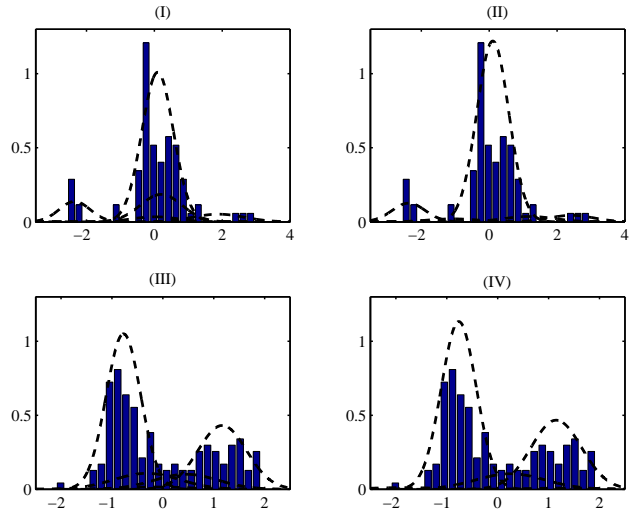}
\caption{Estimated clusters under galaxy data for non-repulsive (I) and repulsive (II) priors and under acidity data for non-repulsive (III)  and repulsive (IV) priors} \label{real:clusters}
\end{figure}

For the galaxy data, figure \ref{data:real} reveals that there are three non-overlapping clusters with the one close to the origin relatively large compared to the others. Although
this large cluster might be interpreted as two highly overlapping clusters, it appears to be well approximated by a single normal density. \cite{reversible} and \cite{escobar}  estimated the number of components, obtaining a posterior distribution on $k$ concentrating on values ranging from $5$ to $7$. This may be due to the non-repulsive prior allowing closely overlapping components, favoring relatively large values of $k$. Figure \ref{real:clusters} reveals that the non-repulsive prior specification leads to two overlapping and essentially indistinguishable clusters.  Under repulsive priors, no clusters overlap significantly and unnecessary components receive a weight close to zero. 
 
\indent For the acidity data, figure \ref{data:real} suggests that two clusters are involved. Since one of them appears to be highly skewed, we expect that three clusters might be needed to approximate this density well. \cite{reversible} obtained a posterior for $k$ almost equally concentrated on values of $k$ ranging from $3$ to $5$. Figure \ref{real:clusters} 
shows the estimated clusters for both repulsive and non-repulsive priors. With non-repulsive priors, four clusters receive significant weight and two of them overlap significantly. With 
repulsive priors, only three clusters receive significant weight and all of them appear fairly separated.

\begin{figure}
\centering
\includegraphics[width=120mm]{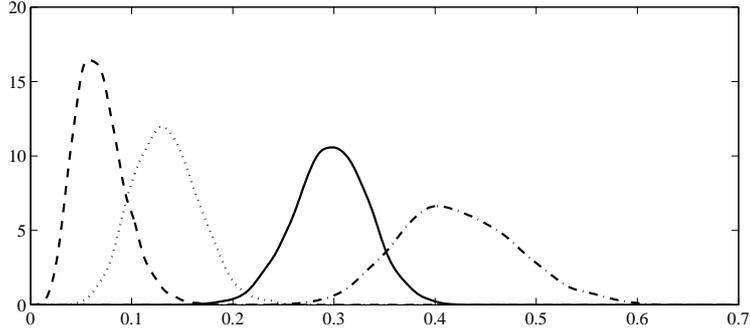}
\caption{Density of sum of extra weights under k=6 for non-repulsive (solid) and repulsive (dash) and k=10 components for non-repulsive (dash-dot) and repulsive (dot)} \label{real:iris}
\end{figure}

The iris data were previously analyzed by \cite{sugar:iris} and \cite{clusterCV} using new methods to estimate the number of clusters based on minimizing loss functions. They concluded the optimal number of clusters was two. This result did not agree with the number of species due to low separation in the data between two of the species.  Such point estimates of the number of clusters do not provide a characterization of uncertainty in clustering in contrast to Bayesian approaches. Repulsive and non-repulsive mixtures were fitted under different choices of upper bound on the number of components. Since the data contains three true biological clusters, with two of these having similar distributions of the available features, we would expect the posterior to concentrate on two or three components. Posterior means and standard deviations of the three highest weights were $(0$$\cdot$$30, 0$$\cdot$$23,0$$\cdot$$13)$ and $(0$$\cdot$$05, 0$$\cdot$$04,0$$\cdot$$04)$ for non-repulsive and $(0$$\cdot$$56, 0$$\cdot$$29,0$$\cdot$$08)$ and $(0$$\cdot$$05, 0$$\cdot$$04,0$$\cdot$$03)$ for repulsive. Clearly, repulsive priors lead to a posterior more concentrated on two components, and assign low probability to more than three components.  Figure \ref{real:iris} shows the density of the total probability assigned to the extra components. This quantity was computed considering the number of species as the true number of clusters. According to figure \ref{real:iris}, our repulsive prior specification leads to extra component weights very close to zero regardless of the upper bound on the number of components. The posterior uncertainty is also small. Non-repulsive mixtures assign large weight to extra components, with posterior uncertainty increasing considerably as the number of components increases. 

\section*{Acknowledgement}
This research was partially supported by a grant from the National Institute of Environmental Health Sciences (NIEHS) of the National Institutes of Health (NIH).
\section*{Supplementary Material}
Supplementary material includes the proof of lemma 2 and lemma 4, assumptions B1-B5, conditions $(i)$, $(ii)$ and $(iii)$ of theorem 3.1. in \cite{Scricciolo2011} and theorem 2.1. in \cite{ghosal2000}.
\appendix 
\section*{Appendix}
\noindent Throughout the appendix we write all constants whose values are of no consequence to be equal to 1.
\begin{proof} [of lemma~\ref{kl}]
By assumption B0, $\vartheta(k=k_0)>0$. We consider the case $f$ is a finite mixture with $k_0$ components. By assumption A1, for each $\eta >0$ there is a corresponding $\delta >0 $ such that, for any given $y\in \mathcal{Y}$ and for all $\gamma_1, \gamma_2 \in \Gamma$ with $|\gamma_1 - \gamma_2|<\delta$, we have that $|\phi(y; \gamma_1) - \phi(y;\gamma_2)| < \eta$.  Let $S_{\delta}=P_{\delta} \times \Gamma_{\delta}$ with $\Gamma_{\delta}=\left\lbrace \gamma : |\gamma_j - \gamma_{0j}| \leq \delta,  j\leq k_{0} \right\rbrace$ and $P_{\delta}=\left\lbrace p :  |p_{j} - p_{0 j}| \leq \delta,  j \leq k_{0} \right\rbrace$. By assumption A1 and A2, for any given $y$ and for any $\eta> 0$, there is a $\delta >0$ such that  $|f_{0} - f| \leq \eta$ if $\theta \in S_{\delta}$. This means that, $f \to f_0$ as $\theta \to \theta_0$, for any given $y$. Equivalently, we can say that  $|\log(f_{0}/f)| \to 0$  pointwise as $\theta \to \theta_0$. Notice that 
\[\left|\log\left(f_{0}/f\right)\right| \leq \left|\log\left\lbrace\sup_{\gamma \in D_{0}} \phi(\gamma)\right\rbrace - \log\left\lbrace\inf_{\gamma \in D_{0}} \phi(\gamma)\right\rbrace\right|\] By assumption A3 and applying the dominated convergence theorem, for any $\epsilon > 0$ there is a $\delta >0$ such that $\int f_0 \log(f_{0}/f) < \epsilon$ if $\theta \in S_{\delta}$.  By the independence of the weights and the parameters of the kernel,\begin{equation*}\Pi(KL(f_0,f)< \epsilon)\geq \lambda( P_{\delta} )\pi(\Gamma_{\delta})\end{equation*}
Assumption A4 combined with the fact that $\{ \gamma: ||\gamma -\gamma_0||_1 \leq \delta\} \subseteq \Gamma_{\delta}$ result in $\pi(\Gamma_{\delta})>0$. Finally, since $\lambda= Dirichlet(\alpha)$, it can be shown that $\lambda( P_{\delta})>0$.\end{proof}
\begin{proof} [of lemma ~\ref{contraction:rate}]
\indent To prove lemma \ref{contraction:rate} we need to show that the three conditions of theorem 2.1 in \cite{ghosal2000} are satisfied. First, define $D(\epsilon, \mathcal{F},d_s)$ as the maximum number of points in $\mathcal{F}$ such that the distance, with respect to metric $d_s$, between each pair is at least $\epsilon$. Let $d_s$ be either the Hellinger metric or the one induced by the L1-norm. For given sequences $k_n, a_n, u_n \uparrow \infty$ and $b_n\downarrow 0$ define $$\mathcal{F}^{(k)}_n=\left\{ f: f=\sum_{j=1}^{k} p_{j} \phi(\mu_{j},\sigma), \mu \in (-a_n,a_n)^k,  \sigma \in (b_n,u_n)\right\}$$ and $\mathcal{F}_n=\cup_{j=1}^{k_n} \mathcal{F}^{(j)}_n$. As it is shown in \cite{Scricciolo2011},  for constants $f_2 \geq f_1 >0$ and $l_1,l_2,l_3>0$, derived below to satisfy condition (2) and (3) in \cite{ghosal2000}, define $f_1 \log n \leq k_n \leq f_2 \log n$, $a_n=l_3\left( \log \bar{\epsilon}_n^{-1}\right)^{1/2}$, $b_n=l_1 (\log \bar{\epsilon}_n^{-1})^{-1/e_2}$ and $u_n=\bar{\epsilon}_n^{-l_2}$, $\log D(\bar{\epsilon}_n,\mathcal{F}_n,d_s)\lesssim n \bar{\epsilon}_n^2$ with $\bar{\epsilon}_n=n^{-1/2} \log n$. 
 
\indent Let $A_{n,j}=(-a_n,a_n)^j$. In order to show condition (2) of theorem 2.1. in \cite{ghosal2000}, we need to show that there is a constant $q_1>0$ such that $\pi(A_{n,k}^C) \lesssim \exp(- q_1 a_n^2)$.  From the exchangeability assumption it follows\\
\begin{tabular}{lllll}
&&$pr(A_{n,k}^C|k=s)$&$=\sum_{j=1}^s \frac{s!}{j! (s-j)!} \pi\left(A_{n,j}^C \times A_{n,s-j}\right)$\\ 
&&&$\leq s \sum_{j=1}^s \frac{(s-1)!}{(j-1)! (s-j)!} \pi\left(A_{n,j}^C\times A_{n,s-j}\right) \leq s \pi_m(A_{n,1}^C)$\\ 
\end{tabular}

Therefore, condition C1 implies that, for a positive constant $q_1$ we have $\pi(A_{n,k}^C)\lesssim E(k)\exp(-q_1 a_n^2)$  with $E(k)<\infty$ by condition (ii) of theorem 3.1. in \cite{Scricciolo2011}. Given a positive constant $z_2$ chosen to satisfy condition (3) in theorem 2.1 of \cite{ghosal2000}, let $f_1 \geq (z_2 + 4)/d_2$, $l_1 \leq \left\lbrace e_1/4 (z_2+4)\right\rbrace^{1/e_2}$, $ l_2 \geq 4(z_2 +4)/e_3$ and $l_3 \geq \left\lbrace4(z_2 +4)/q_1\right\rbrace^{1/2}$. Under these values of $f_1, l_1, l_2$ and $l_3$, following \cite{Scricciolo2011}, assumptions (i), (ii) of theorem 3.1. in \cite{Scricciolo2011} combined with assumption C1 imply $\Pi(\mathcal{F}\setminus \mathcal{F}_n) \lesssim \exp\left\lbrace-(z_2 + 4)n \tilde{\epsilon}_n^2\right\rbrace$ with $\tilde{\epsilon}_n=n^{-1/2}(\log n)^{1/2}$.

\indent To show condition (3) of theorem 2.1 in \cite{ghosal2000}, we can again follow the proof of theorem 3.1. in \cite{Scricciolo2011}. The only thing we need to show is that,  there are constants $u_1, u_2, u_3>0$ such that for any $\epsilon_n \leq u_3$ \[\pi(||\mu-\mu_0 ||_1\leq \epsilon_n) \geq u_1 \exp\left\lbrace-u_2 k_0 \log(1/\epsilon_n)\right\rbrace\] 
that is guaranteed by condition C2.
Therefore, it can be easily showed that, for sufficiently large $n$, $z_2>0$ and $\tilde{\epsilon}_n=n^{-1/2}(\log n)^{1/2}$, $\Pi\left\lbrace B_{KL}(f_0,\tilde{\epsilon}_n^2) \right\rbrace \gtrsim  \exp(-z_2 n \tilde{\epsilon}_n^2)$. \end{proof}
\begin{proof}[of theorem ~\ref{weights:theorem}]  Only for this proof and for ease of notation the density $f$ will be referred as $f_{\theta}$. Define the non identifiability set as $T=\{\theta: f_{\theta} = f_{0} \}$. In order to define each vector in $T$, let $0 = t_0 < t_1< t_2 \ldots < t_{k_0} \leq k$ and $\gamma_j = \gamma_{0i}$ for $j \in I_i=\{t_{i-1}+1, t_i \}$. Let $p_{0i}=\sum_{j=t_{i-1}+1}^{t_i}p_j$ and $p_{j}=0$ for  $j>t_{k_0}$. Define $q_j=p_j/p_{0i}$ for $j \in I_i$. \\
 \indent Define $A_n=\left\lbrace \min_{\{\iota \in S_k\}}\left( \sum_{i=1}^{k-k_0} p_{\iota(i)}\right) > \delta_n M_n \right\rbrace$ and  $A'_n=A_n \cap \{ \|f-f_0\|_1\leq M \delta_n \}$. Let  $D_n= \int_{ \{ \|f-f_0\|_1<\delta_n \}} \exp(l_n(\theta) - l_n(\theta_0)) d(\pi \times \lambda)(\theta)$ with $l_n(\theta_0)$ being the log-likelihood evaluated at $\theta_0$. Along the line of \cite{Rousseau}'s proof, to prove theorem 1 we need to show that for any $\epsilon>0$ there are positive constants $m_1, m_2$ and a permutation $\iota \in S_k$ such that\\
 \begin{minipage}{0.5\linewidth} \begin{equation}D_n \geq m_1 n^{-s(k_0,\alpha)/2}  \label{condDn}\end{equation}
\end{minipage} 
\begin{minipage}{0.5\linewidth} 
\begin{equation} \Pi(A_n')\leq m_2 \delta_n^{s(k_0,\alpha)} M_n^{\bar{\alpha}-m/2}   \label{condAn}\end{equation}\end{minipage}\\
 
 with $s(k_0,\alpha)=k_0-1+m k_0+\sum_{j=1}^{k-k_0}\alpha_{\iota(j)}$.  Following \cite{Rousseau}'s proof, we can show that, under condition B5, (\ref{condDn}) is satisfied for sufficiently large $n$. Concerning (\ref{condAn}), \cite{Rousseau} showed that on $A_n'$, there is a set $I_i$ containing indices $j_1$ and $ j_2$ such that
   \[|\gamma_{j_1}-\gamma_{0i}| \leq \left(\delta_n/q_{j_1}\right)^{1/2}, \quad |\gamma_{j_2}-\gamma_{0i}|\leq \left(\delta_n/q_{j_2}\right)^{1/2}\] with $q_{j_1}>\epsilon/k_0$ and $q_{j_2}>\delta_n M_n/2$. Therefore, from the triangle inequality it follows \[|\gamma_{j_1}-\gamma_{j_2}|\leq \left\lbrace2\delta_n/\min(q_{j_1},q_{j_2})\right\rbrace^{1/2}\] Now, for sufficiently large $n$, $\min(q_{j_1},q_{j_2})> \delta_n M_n/2$ and therefore $|\gamma_{j_1}-\gamma_{j_2}|\leq M_n^{-1/2}$. Recalling that $g$ is bounded above by a positive constant, there exists a constant $c>0$ such that
\begin{equation}h(\gamma)\leq c g\left\lbrace d(\gamma_{j_1},\gamma_{j_2})\right\rbrace \leq c g\left(M_n^{-1/2}\right)\label{bound}\end{equation}
\noindent
Let the prior probability of the set $A'_n$ be defined as $\Pi(A'_n) = \int_{A'_n}  d(\pi \times \lambda)(\gamma \times p)$. To find an upper bound for this integral, directly apply the proof of \cite{Rousseau} showing that  $\Pi(A'_n)\leq g\left(M_n^{-1/2}\right)\delta_n^{s(k_0,\alpha)} M_n^{\bar{\alpha} - m/2}$. By assumption, for sufficiently large $n$, $g\left(M_n^{-1/2}\right)\leq r_1 M_n^{-r_2}$. Letting $s_{r_2}=r_2+m/2-\bar{\alpha}$, it follows 
\[\Pi(A'_n)\leq M_n^{-s_{r_2}} (\log n)^{q s(k_0,\alpha)} D_n\] Therefore, $M_n=(\log n)^{q s(k_0,\alpha)/s_{r_2}}$ implies $\Pi(A'_n)=O_p(D_n)$.\end{proof}

\bibliographystyle{biometrika} 
\bibliography{repulsive_mixtures_rev_arXiv} 

\end{document}